\documentclass[twocolumn,showpacs,preprintnumbers,amsmath,amssymb,aps]{revtex4}
\usepackage{graphicx}
\usepackage{dcolumn}
\usepackage{bm}

\begin{document}

\title{Effect of Polymerization on the Boson Peak, from the Liquid to Glassy
Phase}

\author{S. Caponi$^{1,2}$, S. Corezzi$^{3,4}$, D.
Fioretto$^{2,4}$, A. Fontana$^ {1,2}$, G. Monaco$^{5}$, F.
Rossi$^{1,2}$} \affiliation{$^1$ Dipartimento di Fisica,
Universit\`{a} di
Trento, Via Sommarive 14, I-38050 Povo (Trento), Italy\\
$^2$CNR-INFM CRS Soft, Universit$\grave{a}$ di Roma ``La
Sapienza", P. A. Moro 2, I-00185 Roma, Italy\\
$^3$CNR-INFM Polylab, Universit$\grave{a}$ di Pisa,
Largo Pontercorvo 3, I-56127 Pisa, Italy\\
$^4$Dipartimento di Fisica, Universit$\grave{a}$ di
Perugia, Via A. Pascoli, I-06100 Perugia, Italy\\
$^5$European Synchrotron Radiation Facility, BP 220, F-38043
Grenoble, France}

\date{\today}

\begin{abstract}
Raman scattering measurements are used to follow the modification of
the vibrational density of states in a reactive epoxy--amine mixture
during isothermal polymerization. Combining with Brillouin light and
inelastic x-ray scattering measurements, we analyze the variations
of the boson peak and of the Debye level while the system passes
from the fluid to a glassy phase upon increasing the number of
covalent bonds among the constituent molecules. We find that the
shift and the intensity decrease of the boson peak are fully
explained by the modification of the elastic medium throughout the
reaction. Surprisingly, bond--induced modifications of the structure
do not affect the relative excess of states over the Debye level.
\end{abstract}


\maketitle

The excess of low--frequency modes over the Debye level in the
vibrational density of states $g(\omega)$ of disordered materials,
the so--called boson peak (BP), remains the object of controversial
debate \cite{TaraskinPRL2001, GrigeraNATURE2003, GotzePRE2000,
GurevichPRB2003, SchmidPRL2008, RuffléPRL2008, BuchenauCM2007,
PillaJPCM2004}. The BP has been investigated on temperature
\cite{FontanaEPL1999, CaponiPRB2007,SteurerPRL2008}, on pressure
\cite{ InamuraPHYSB2000, MonacoPRL2006den, NissPRL2007,
SugaiPRL1996}, and on changes in sample preparation
\cite{MonacoPRL2006hyp, KojimaPB1999, CaponiJPCM2007} by
measurements of specific heat, Raman, neutron, and nuclear inelastic
scattering. Sample densification is common to all the conditions
where significant changes of the BP have been observed. In this
respect, the various techniques reveal the same trend: an increase
in density causes a decrease in the BP intensity and a shift of its
maximum toward higher frequencies. A so clear experimental evidence,
however, does not correspond to an as clear interpretation.

Recent studies suggest the central role of changes of elastic properties in the BP evolution, although general
consensus has not emerged: shift and intensity decrease of the BP for hyperquenched \cite{MonacoPRL2006hyp} and
permanently densified glasses \cite{MonacoPRL2006den} can be fully explained by the corresponding changes of the
Debye level, which describes the elastic medium; on the other hand, for a polymer glass the effect of pressure
on the BP is found to be stronger than elastic medium transformation, at least for exceptionally high levels of
densification \cite{NissPRL2007}.
 Following the idea in
\cite{MonacoPRL2006hyp}, any variable which defines the Debye level
(not only density, but also sound velocity) could play a decisive
role in the BP modification. Despite the effects of sound velocity
may be very pronounced in the liquid
---where the BP continues to be present \cite{Kalampounias,BrodinPRB1996}---
studies are limited to glass--formers below their $T_{g}$, and only temperature and pressure--induced effects
have been explored. An even more intriguing aspect concerns the limiting effect, for the BP scaling properties,
of changes in the local structure. Results for permanently densified glasses indeed suggest that these
properties may be lost in the presence of structural modifications in the short--range order
\cite{MonacoPRL2006den}.

In this Letter we consider effects on the BP and on the Debye level in a reactive mixture as the monomers,
initially liquid, irreversibly polymerize under constant $T$ and $P$. As reaction proceeds, an increasing number
of loose van der Waals bonds are replaced by stiffer covalent bonds that slowdown the molecular motions and
ultimately lead to a frozen, glassy structure (chemical vitrification \cite{CorezziNATURE2002,CorezziPRL2005}).
Meanwhile, the sample density and sound velocity increase, both contributing to significantly change the Debye
level. Given the nature of the process, chemical vitrification double challenges the low--frequency vibrational
properties to scale with elastic properties. First, it is not obvious that the vibrational density of states
scales in the liquid the same way as in the glass. Second, the gradual passage from the fluid to a glassy phase
is clearly accompanied by a large number of microscopic structural changes
---i.e., local reorganization of atoms in different molecular
groups--- carried out by the bonding process.

Starting from the initially liquid mixture, we have monitored by Raman scattering (RS) the density of
vibrational states and measured by Brillouin light (BLS) and inelastic x--ray scattering (IXS) the sound
velocities and elastic moduli, throughout the reaction. Our results show that, even in the liquid phase, the
transformation of the elastic medium continues to describe the BP evolution. Unexpectedly, bond--induced
modifications of the structure that affect both atomic vibrations and slow cooperative motions, do not affect
the relative excess of states over the Debye level.

The system we study is an epoxy--amine mixture consisting of
diglycidyl ether of bisphenol--A (DGEBA) and diethylenetriamine
(DETA) in the stoichiometric ratio of 5:2, during polymerization at
a constant $T$=275.3 K. At this temperature the monomers bond slowly
to each other and the total reaction time is $\sim 2$ days. The
reaction proceeds by stepwise addition of the amino hydrogen to the
epoxy group, without elimination of by--product molecules. Starting
from a liquid sample ($T_{g}\sim 230$ K), the final product is a
glassy network--polymer. In all experiments, the reagents were mixed
in a glass container, stirred for about 2 min, and then transferred
in the measurement cell (a cylindrical pyrex cell of inner diameter
10 mm for RS and BLS measurements, a stainless steel cell equipped
with kapton windows for IXS measurements). The RS experiment was
performed using a standard experimental setup Jobin Yvon U1000, in a
wide frequency range (from 3 to 4500 cm$^{-1}$) in order to follow
also the evolution of the molecular peaks. The light polarizations
were perpendicular (VV) and orthogonal (HV) to the scattering plane.
The IXS experiment was performed at the beam line ID16 of the
European Synchrotron Radiation Facility with an incident photon
energy of 21.748 keV and an instrumental energy resolution of 1.5
meV \cite{CorezziPRL2006}. Spectra at different values of the
reaction time were taken in a $Q$ range from 1 to 10 nm$^{-1}$. BLS
unpolarized (VU) spectra were measured in $90^{\circ}$--scattering
geometry using a tandem Fabry--Perot interferometer. The
measurements were performed with a laser light of 532 nm wavelength.
The refractive index at the same wavelength was measured throughout
reaction by the angle of minimum deflection. The sample density
$\rho$ as a function of time was obtained as reported in
Ref.~\onlinecite{Cauchy-unpublished}; the whole variation is
$\Delta\rho \sim 4\%$.

\begin{figure}
\vspace{-0.3 cm} \hspace{-0.5 cm}
\includegraphics[width=0.50\textwidth]{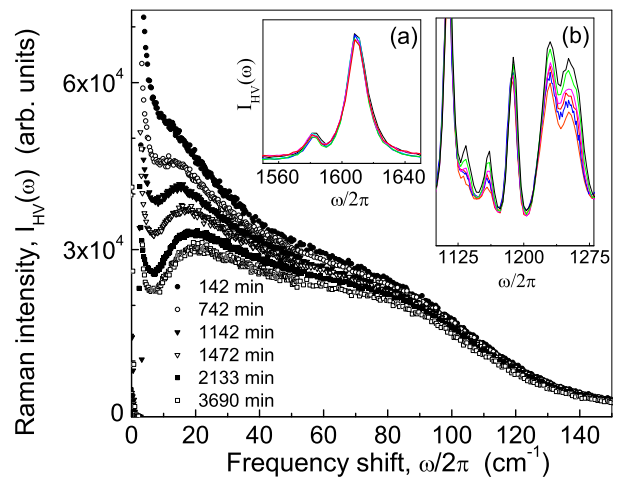}
\vspace{-0.8 cm} \caption {\label{Fig:spectraRAW} Depolarized Raman
spectra at some selected reaction times as indicated in the legend,
during the isothermal reaction DGEBA-DETA 5:2 at $T$=275.3 K. The
insets show the same spectra as in the main panel, (a) in the
frequency range of the molecular peak at 1610 cm$^{-1}$ (aromatic
quadrant stretch) invariant on reaction, and (b) in the frequency
range of the band at 1260 cm$^{-1}$ (breathing of the epoxide ring),
whose intensity decrease corresponds to the epoxy group
consumption.}
\end{figure}

Figure~\ref{Fig:spectraRAW} shows the experimental depolarized Raman
intensity, $I_{HV}(\omega)$. The spectra are collected sequentially
during the reaction; they do not present any luminescence background
and the data treatment does not require any intensity normalization
factor [inset (a) of Fig.~\ref{Fig:spectraRAW}]. Intensity
variations in the high frequency region only occur for those peaks
related to vibrations of atoms involved in the reaction [inset (b)
of Fig.~\ref{Fig:spectraRAW}]. The depolarization ratio, i.e., the
ratio of depolarized [$I_{HV}(\omega)$] to polarized
[$I_{VV}(\omega)$] Raman intensity, is constant (from 5 to 80
cm$^{-1}$) for all the spectra, showing that no depolarization
mechanism is activated by the chemical changes. It is clear the
presence of the boson peak, which contributes to the spectra with a
maximum at $\sim 15$ cm$^{-1}$, and of the quasielastic scattering
(QES), centered at zero frequency. As reaction proceeds, the QES
significantly decreases till it becomes negligible, and the BP
shifts toward higher frequencies while its intensity decreases, in
agreement with previously reported observations on changing
thermodynamic variables \cite{BrodinPRB1996}.

\begin{figure}
\vspace{-0.8 cm}
\includegraphics[width=0.50\textwidth]{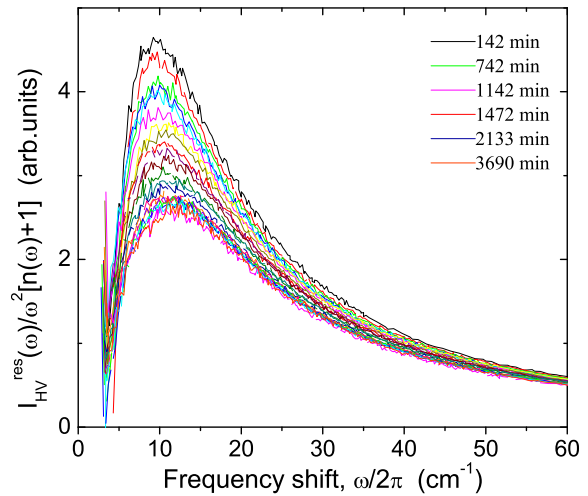}
\vspace{-0.8 cm} \caption{\label{Fig:gWsuW2} The quantity
$I_{HV}^{res}(\omega)/\{\omega^2[n(\omega)+1]\}$
---proportional to the reduced density of
vibrational states, $g(\omega)/\omega^{2}$--- obtained from the
Raman spectra after subtraction of the QES contribution, at
different reaction times.}
\end{figure}

As suggested in the literature \cite{SokolovEPL1997,
WiedersichPRB2001, KojimaPRB1996, WinterlingPRB1975,
PhillipsRPP1987} QES is usually ascribed to some kind of
relaxational process. Therefore, we effectively model the Raman
intensity $I_{HV}(\omega)$ in the frequency region of interest (up
to $\sim 60$ cm$^{-1}$) as a superposition of two contributions, one
($I_{QES}$) related to relaxational processes, and the other
($I_{BP}$) related to the density of vibrational states $g(\omega)$,
i.e.
\begin{eqnarray}
I_{HV}(\omega)&=&I_{QES}(\omega)+I_{BP}(\omega)\nonumber \\
   &=&C^{rel}(\omega)L(\omega)+C^{vib}(\omega)g(\omega)[n(\omega)+1]/\omega~,
\label{eq:I-HV}
\end{eqnarray}
where $C^{rel}(\omega)$ and $C^{vib}(\omega)$ are the vibration--to--light coupling coefficients for the
relaxational and vibrational term, and $n(\omega)+1=\{1-\exp(-\hbar \omega/k_{B}T)\}^{-1}$ is the Bose
population factor. Since in this work we are only interested in the BP variations, our analysis aims at
subtracting from the total signal the QES contribution, described by the function $L(\omega)$. Here, we
approximate $I_{QES}(\omega)$ with an effective Lorentzian shape centered at zero frequency \cite{CaponiPRB2007,
yannopulosPRE2001}, and adopt the criterion to keep the width of this line fixed for all the spectra and to
adjust the intensity in such a way that in the subtracted spectra the QES contribution is totally cancelled.
This operation is consistent with assuming that $C^{rel}(\omega)$ at frequency lower than the BP region is
almost constant, in agreement with experimental determinations made in other systems \cite{FontanaEPL1999,
CaponiPRB2007, SokolovPRB1995}. The residual spectra, $I_{HV}^{res}(\omega)$, only represent the second term in
the rhs of Eq.~(\ref{eq:I-HV}). Using $C^{vib}(\omega)\sim\omega$ for frequencies near the BP maximum
\cite{Nota}, we thus obtain a quantity proportional to the reduced density of vibrational states
$g(\omega)/\omega^{2}$, as
\begin{equation}
\frac{I_{HV}^{res}(\omega)}{\omega^2[n(\omega)+1]} \propto
\frac{g(\omega)}{\omega^{2}}~, \label{eq:gW}
\end{equation}
presented in Fig.~\ref{Fig:gWsuW2}. Here the BP is clearly visible for $\omega >5$ cm$^{-1}$, undergoing a
variation during the reaction of $\sim 50\%$ in intensity and $\sim 20\%$ in frequency position.

\begin{figure}
\vspace{-0.6 cm}
\includegraphics[width=0.50\textwidth]{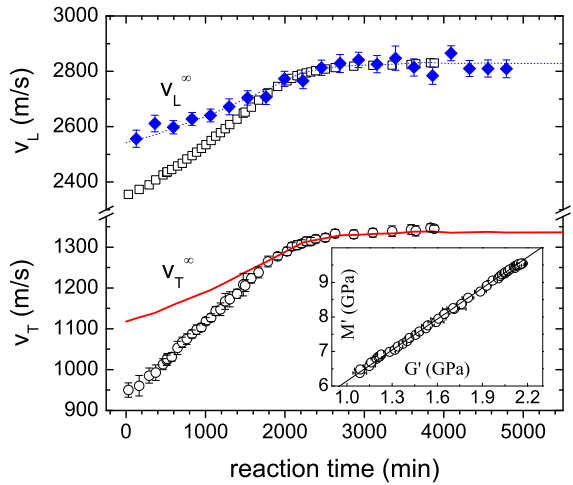}
\vspace{-0.8 cm} \caption{\label{Fig:sound} Longitudinal $v_{L}$
($\square$) and transverse $v_{T}$ ($\circ$) sound velocities
measured by BLS, high--frequency unrelaxed value of the longitudinal
sound velocity ${v_{L}}^{\infty}$ ($\blacklozenge$) measured by IXS,
and high--frequency unrelaxed value of the transverse sound velocity
${v_{T}}^{\infty}$ (solid line) obtained using the Cauchy--like
relation shown in the inset. Inset: Apparent moduli measured by BLS.
The linear fit gives $M' = (3.16 \pm 0.08) + (2.99 \pm 0.02) G'$.}
\end{figure}

\begin{figure}
\vspace{-0.6 cm}
\includegraphics[width=0.49\textwidth]{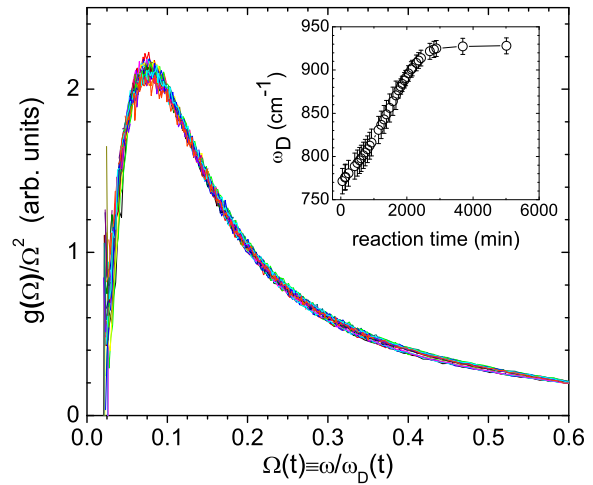}
\vspace{-0.8 cm} \caption{\label{Fig:rescale} Reduced density of
vibrational states after rescaling by the Debye frequency $\omega
_{D}$, at different times of reaction. Data are the same as in
Fig.~\ref{Fig:gWsuW2}. The thickness of the master curve compares
with the uncertainty in the $x$-- and $y$--scaling factors. Inset:
$\omega _{D}$ as a function of reaction time.}
\end{figure}

We now compare these variations with the variations of the Debye
frequency, $\omega _{D}$, given by
\begin{equation} {\omega _{D}}^{3}=\frac{6 \pi^{2} \rho N_{A}N_{F}\langle
v\rangle^{3}}{M} \label{eq:DebyeEnergy}
\end{equation}
where $\rho$ is the density, $N_{A}$ the Avogadro's number, $N_{F}$ the number of atoms per molecule, $M$ the
molar weight, and $\langle v\rangle$ the mean sound velocity defined as
\begin{equation}
\frac{1}{\langle
v\rangle^{3}}=\frac{1}{3}(\frac{1}{v_{L}^{3}}+\frac{2}{v_{T}^{3}})
\label{eq:velocity}
\end{equation}
with $v_{L}$ and $v_{T}$ the longitudinal and transverse sound
velocities. As the number of atoms in the mixture is unchanged on
reaction (no molecule is expelled as by--product), $N_{F}$ and $M$
for our sample were calculated by averaging the corresponding values
for the two components (DGEBA and DETA) weighted by their molar
fraction.

It has to be noticed that the $v_{L}$ and $v_{T}$ values in Eq.~(\ref{eq:velocity}) are those corresponding to
the frequencies of the BP ($\sim 0.5$ THz), therefore the high--frequency unrelaxed values ${v_{L}}^{\infty}$
and ${v_{T}}^{\infty}$. For the studied process these values do not coincide, but in the glass, with the sound
velocities measured by BLS. The results for ${v_{L}}^{\infty}$, obtained by fitting the IXS data at $Q=1$ and 2
nm$^{-1}$ using the damped harmonic oscillator model, are shown in Fig.~\ref{Fig:sound}. The IXS data agree well
with the longitudinal sound velocity measured by BLS in the final stage of reaction, i.e., close and below the
glass transition. However, at an early stage of reaction the IXS velocity is up to 10$\%$ higher than the BLS
one. This difference reveals the effect on acoustic modes of the structural relaxation. On approaching the
glassy state, the relaxation active in the fluid shifts toward lower frequencies and the apparent velocity
$v_{L}$ measured by BLS in the GHz domain tends to its unrelaxed value ${v_{L}}^{\infty}$, always measured by
IXS.

To evaluate ${v_{T}}^{\infty}$ we exploit the Cauchy--like relation
$M' = a + b G'$, between the real part of the longitudinal $M'=\rho
v_{L}^{2}$ and transverse $G'=\rho v_{T}^{2}$ elastic moduli, which
is found to hold true, with the same parameters, for the unrelaxed
moduli and for the apparent moduli measured at finite frequencies
throughout reaction \cite{Cauchy-unpublished} (inset of
Fig.~\ref{Fig:sound}). For DGEBA-DETA 5:2 the parameters are $a=3.16
\pm 0.08$ and $b=2.99 \pm 0.02$. The obtained ${v_{T}}^{\infty}$ is
reported in Fig.~\ref{Fig:sound}. It contributes, together with
${v_{L}}^{\infty}$, to an $\sim 19\%$ increase of $\langle v\rangle$
from the beginning to the end of the reaction.

The calculated Debye frequency $\omega _{D}$ (inset of Fig.~\ref{Fig:rescale}) increases on reaction by $\sim
20\%$. In order to remove this effect from the transformation of $g(\omega)/\omega^{2}$ shown in
Fig.~\ref{Fig:gWsuW2}, we measure the frequencies in Debye frequency units, $\Omega(t)=\omega/\omega_{D}(t)$,
while imposing the conservation of the total number of vibrational states, $g(\Omega)d\Omega=g(\omega)d\omega$.
The rescaled spectra $g(\Omega)/\Omega^{2}$ are calculated as
\begin{equation}
\frac{g(\Omega)}{\Omega^{2}}=\frac{g(\omega)}{\omega^{2}}\omega_{D}^3(t)~.
\label{eq:rescale}
\end{equation}

Within the experimental uncertainty in the Debye frequency
[$\Delta\omega_{D}(t)\sim 1.9\%$ for $t<1700$ min, $\sim 1\%$ for
$t>1700$ min] the obtained rescaled spectra collapse one on the
other without any adjusting parameter (Fig.~\ref{Fig:rescale}).
Thus, the BP variations on polymerization are fully described by the
transformation of the elastic continuum. They hide no appreciable
change in the relative excess of states above the Debye level,
$3/\omega_{D}^{3}$: the decrease of vibrational states is
compensated by the corresponding decrease of the level.

Our results are in agreement with studies of hyperquenched and permanently densified glasses
\cite{MonacoPRL2006hyp, MonacoPRL2006den}, for a shift in the BP position even more pronounced. However, two
important aspects of the reactive process characterize the present with respect to previous investigations.
First, on increasing the number of covalent bonds the system passes from the fluid to a glassy phase; the
scaling property of the BP with the elastic continuum deformation holds throughout the reaction, i.e., above as
well as below the glass transition. In this respect, it should be noted that, in contrast to the previously
studied cases, most of the Debye level variation is due to the sound velocity increase that occurs above the
glass transition, rather than to sample densification, suggesting that density not necessarily plays a dominant
role in the BP transformation. Second, on polymerization, changes of the macroscopic properties are not induced
by thermodynamic variables; instead, they are driven by changes of a chemical nature, which also modify the
structure of the sample on a mesoscopic and microscopic length scale. As the number of chemical bonds increases,
modifications of the intermediate--range order appear in the x--ray scattering data as a progressively
increasing pre--peak at small wave vectors in the static structure factor \cite{CorezziPRL2006}; modifications
of the local environment around the atoms involved in the reaction are revealed by the high--frequency Raman
scattering data, as shown in Fig.~\ref{Fig:spectraRAW}(b). These continuous changes in the structure seem not to
affect the relative distribution of excess vibrational states over the Debye level.

In conclusion, we studied the variations of the vibrational spectra induced by covalent bond formation during
the process of chemical vitrification of a reactive mixture. We find that the BP, as the polymer structure
develops, always scales with the corresponding value of the Debye frequency. The results of this study indicate
that a significant transformation of the BP --- even in the liquid phase --- is sufficiently described by the
changes of the elastic properties of the medium, independently of the variable inducing those changes.

\end{document}